\documentstyle{article}
\begin{document}
\newcommand{\norm} [1] {\parallel \hspace{-0.15cm} #1
\hspace{-0.15cm} \parallel}
\newcommand{\betrag} [1] {\mid \hspace{-0.13cm} #1
\hspace{-0.15cm} \mid}
\newcommand{\rsta} [1] {\mid \hspace{-0.1cm} #1 \hspace{-0.04cm}
\rangle}
\newcommand{\lsta} [1] {\langle \hspace{-0.04cm} #1
\hspace{-0.12cm} \mid}
\begin{titlepage}
\centerline{\normalsize DESY 93 -145 \hfill ISSN 0418 - 9833}
\centerline{\hfill hep-th/9311018}
\centerline{\normalsize October 1993 \hfill}
\vskip.6in
\begin{center}
{\Large Sum-over-histories representation for the causal Green
function of free scalar field theory}
\vskip.6in
{\Large Oliver Rudolph $^*$}
\vskip.3in
{\normalsize \em II. Institut f\"ur Theoretische Physik, Universit\"at
Hamburg}
\vskip.05in
{\normalsize \em Luruper Chaussee 149}
\vskip.05in
{\normalsize \em 22761 Hamburg, Germany}
\end{center}
\normalsize
\vfill
\begin{center}
{ABSTRACT}
\end{center}
\smallskip
\noindent
A set of Green functions ${\cal G}_{\alpha}(x-y), \alpha \in
[0, 2 \pi [$, for free scalar field theory is introduced, varying
between the Hadamard Green function $\Delta_1(x-y) \equiv
\linebreak[2] \lsta{0} \hspace{-0.1cm}
\{ \varphi(x), \varphi(y) \} \hspace{-0.1cm} \rsta{0}$ and the
causal Green function  ${\cal G}_{\pi}(x-y) = i \Delta(x-y) \equiv
[\varphi(x), \varphi(y)]$.
For every $\alpha \in [0, 2 \pi [$ a path-integral representation
for ${\cal G}_{\alpha}$ is obtained both in the configuration
space and
in the phase space of the classical relativistic particle.
Especially setting $\alpha = \pi$ a sum-over-histories
representation for the causal Green function is obtained.
Furthermore using BRST theory an alternative path-integral
representation for ${\cal G}_{\alpha}$ is presented. From these
path integral representations the composition laws for the ${\cal
G}_{\alpha}$'s are derived using a modified path decomposition
expansion.
\bigskip \noindent
\centerline{\vrule height0.25pt depth0.25pt width4cm \hfill}
\noindent
{\footnotesize $^*$ Supported by Deutsche Forschungsgemeinschaft}
\end{titlepage}
\newpage
\section{Introduction}
Relativistic quantum mechanics of a spinless point particle is a
simple example of a system with constraints and is often viewed as
a toy-model for more complicated field theories possessing gauge
symmetry or reparametrization invariance, e.g. quantum cosmology.
\\
With the aim of understanding better the sum-over-histories
formulation of quantum cosmology Halliwell and Ortiz  considered
in an interesting recent paper \cite{Ha92}
sum-over-histories/path-integral representations of various Green
functions
of the Klein-Gordon equation, constructed canonical
representations and investigated the relationship between these
representations and the existence of composition laws for these
Green functions.
A canonical representation of a Green function is a
representation as an inner product ${\cal G}(x''\mid x') =
\lsta{x''}  x' \rangle$, where $\rsta{x}$ denotes a complete set
of configuration space states for any fixed $x^0$.
A sum-over-histories representation is a representation of $\cal
G$ of the
form ${\cal G}(x'' \mid x') = \sum \exp(i S)$, where the sum runs
over all possible paths from $x'$ to $x''$ and $S$ denotes the
action of each path. \\
However Halliwell and Ortiz failed to obtain a path integral and
sum-over-histories representation for the causal Green function.
The purpose of this paper is to fill this gap. \\
This paper deals exclusively with the case of a spinless
relativistic point particle. The reader
interested in the connection to quantum cosmology is referred to
the References, e.g. \cite{Ha92,Ha88,Te82}. \\
The paper is organized as follows:
In the Introduction some results of certain Green functions in
free scalar
field theory, relativistic classical mechanics of a point
particle and relativistic quantum mechanics of a point
particle are shortly reviewed for convenience. For complete
details the reader is once again referred to the References.
A set of Green functions ${\cal G}_{\alpha}(x-y), \alpha \in
[0, 2 \pi [$, varying
between the Hadamard Green Function and the causal Green function
is introduced.
Several path integral representations in phase and in
configuration space for the ${\cal G}_{\alpha}$'s
are presented in the second section. In the third section a
modified path decomposition expansion is presented and used to
derive the relativistic composition laws for the ${\cal
G}_{\alpha}$'s
from their sum-over-histories representations.
\subsubsection*{Green functions for free scalar field theory}
The {\em causal Green function} of free scalar field theory is
defined by \footnote{The notations in this work differ from those
in
\cite{Ha92}.} \begin{equation} i \Delta(x-y) \equiv [\varphi(x),
\varphi(y)]. \end{equation}
$\Delta$ is Lorentz invariant and vanishes outside the light-cone.
The following identities are well known
\begin{eqnarray} \label{x} i \Delta(x-y) & = & \frac{1}{(2 \pi)^3}
\int d^4k \epsilon(k_0) \delta(k^2-m^2) e^{-ik(x-y)} \\
\Delta(x-y) & = & \frac{-1}{(2 \pi)^3} \int \frac{d^3 {\bf
k}}{\omega_{\bf \small k}} \sin(\omega_{\bf \small k}(x^0 -y^0))
e^{i {\bf \small k}({\bf \small x} - {\bf \small y})},
\end{eqnarray}
where \begin{equation} \label{g} \epsilon(k_0) = \left\{
\begin{array}{r@{\quad:\quad}l} 1 & k_0 > 0 \\ -1 & k_0 < 0
\end{array} \right. \end{equation}
and \begin{equation} \label{om} \omega_{\bf \small k} \equiv +
\sqrt{m^2 + \sum_{i=1}^3 k_i^2}. \end{equation}
The {\em Hadamard Green function} of free scalar field theory is
defined as \begin{equation} \Delta_1(x-y) \equiv
\lsta{0} \hspace{-0.1cm} \{
\varphi(x), \varphi(y) \} \hspace{-0.1cm} \rsta{0}. \end{equation}
The Hadamard Green function may be written
\begin{eqnarray} \Delta_1(x-y) & = & \frac{1}{(2 \pi)^3} \int d^4k
\delta(k^2-m^2) e^{-ik(x-y)} \\
& = & \frac{1}{(2 \pi)^4} \int_{-\infty}^{\infty} dT \int d^4k
e^{-ik(x-y) +iT(k^2-m^2) } \\
\label{y} & = & \frac{1}{(2 \pi)^3} \int \frac{d^3 {\bf
k}}{\omega_{\bf \small k}} \cos(\omega_{\bf \small k}(x^0 -y^0))
e^{i {\bf \small k}({\bf \small x} - {\bf \small y})}.
\end{eqnarray}
Both the causal and the Hadamard Green function are solutions of
the Klein-Gordon equation $(\Box + m^2) \varphi =0$.
\subsubsection*{Relativistic point particle}
A point particle in Minkowski spacetime follows a worldline
$x(\tau)$, which is parametrized by a variable $\tau$ - not
necessarily time -, which is monotonically increasing along the
wordline, whether the worldline is proceeding forward or backward
in time.
The classical action is given by \begin{equation} S_0[x(\tau)] =
\int_{\tau'}^{\tau''} d \tau L, \end{equation}
where \begin{equation} L = -m \left[ \eta_{\mu \nu}  \dot{x}^{\mu}
\dot{x}^{\nu} \right]^{\frac{1}{2}} = -m \sqrt{\dot{x}^2}.
\end{equation}
The dot denotes differentiation with respect to $\tau$ and
$\eta_{\mu \nu}$ is the usual diagonal metric in Minkowski space
with signature ($+,-,-,-$). The action $S_0$ is invariant under
reparametrisations $\tau \mapsto f(\tau)$, provided $f$ satisfies
$\frac{df}{d \tau} > 0$, $f(\tau')= \tau'$ and $f(\tau'') =
\tau''$. \\
Introducing the momenta conjugate to $x^{\mu}(\tau)$
\begin{equation} p_{\mu} = \frac{\partial L}{\partial
\dot{x}^{\mu}} = -m \frac{\dot{x}_{\mu}}{\sqrt{\dot{x}^2}},
\end{equation}
one finds immediately the well known first-class primary
constraint
\begin{equation} \label{a} {\cal H}_0 \equiv p_{\mu}p^{\mu} -m^2
=0. \end{equation}
One also finds that the canonical Hamiltonian $H = p_{\mu}
\dot{x}^{\mu} - L$ vanishes, which is typical for
reparametrization invariant systems.
\footnote{Actually the following is true: if coordinates and
momenta for a generally covariant system transform under
reparametrisations as scalars, then the Hamiltonian vanishes
\cite{He92}.}
The constraint (\ref{a}) can incorporated into the Hamiltonian
action via a Lagrange multiplier $N(\tau)$ by writing
\begin{equation} \label{b} S = \int_{\tau'}^{\tau''} d \tau \left(
p_{\mu} \dot{x}^{\mu} - N {\cal H}_0 \right). \end{equation}
The classical equations of motion are obtained
by extremizing $S$ under independent variations of $x, p$ and $N$
subject to the conditions $x(\tau')=x'$ and $x(\tau'')=x''$. \\
The action (\ref{b}) is still invariant under the following gauge
transformations generated by ${\cal H}_0$
\begin{equation} \label{c} \delta x = \epsilon(\tau) \{x, {\cal
H}_0\}, \hspace{0.5cm} \delta p = \epsilon(\tau) \{p, {\cal
H}_0\}, \hspace{0.5cm} \delta N= \dot{\epsilon}(\tau),
\end{equation}
where $\epsilon(\tau)$ is an arbitrary infinitesimal function
satisfying
$\epsilon(\tau')=\epsilon(\tau'') =0$ and $\{ \cdot , \cdot \}$
denotes the Poisson-bracket. \\
In order to fix the
reparametrization invariance (\ref{c}) one has to impose a further
restriction (gauge-fixing condition).
In \cite{Te82} it has been shown that the simplest gauge-fixing
condition, which fully fixes the gauge and does not remove
physical
modes from the system is \begin{equation} \label{d} \dot{N}=0.
\end{equation}
This so-called {\em proper time gauge} is adopted throughout this
paper. More generally gauge conditions of the form $\dot{N} =
\chi(p,x,N)$ are also admissible. Canonical gauge conditions of
the
form $C(p,x) =0$ for all $\tau$ would in general violate
eq.~(\ref{c}).
In \cite{Te82} it has also been pointed out that in the
gauge (\ref{d}) the transition amplitude from $x'$ to $x''$ is
given by
\begin{equation} \label{e} \Delta_F(x'' \mid x') = - i
\int^0_{-\infty}
dN(\tau''-\tau') \int {\cal D}p {\cal D}x \exp \left( i
\int_{\tau'}^{\tau''} d \tau (p \dot{x} -N {\cal H}_0) \right),
\end{equation}
where $\Delta_F(x'' \mid x') \equiv -i \lsta{0} \hspace{-0.1cm}
T(\varphi(x'') \varphi(x')) \hspace{-0.1cm} \rsta{0}$ denotes the
Feynman propagator of free scalar field theory. \\
On the other hand in \cite{Ha88} it has been shown that
eq.~(\ref{e}) yields $\Delta_1$, provided the range of $N(\tau'' -
\tau')$ is
taken to be $-\infty$ to $\infty$
\begin{equation} \label{f} \Delta_1(x'' \mid x') = \int_{-
\infty}^{\infty} dN(\tau''-\tau') \int {\cal D}p {\cal D}x \exp
\left( i \int_{\tau'}^{\tau''} d \tau (p \dot{x} -N {\cal H}_0)
\right). \end{equation}
As usual the path integrals (\ref{e}),(\ref{f}) are thought to be
defined by a time slicing procedure with the Liouville measure on
each time-slice and the following boundary conditions at the end
points
\begin{equation} \label{w} x(\tau') = x'; x(\tau'') = x''.
\end{equation}
In eq.~(\ref{r}) below the time slice definition for a path
integral is explicitly given. \\
An expression of the form (\ref{f}) is formally derived in
\cite{Ha88} using BRST symmetry. \\
Recalling that \begin{eqnarray} \Delta_1(x-y) & = & \lsta{0}
\hspace{-0.1cm} \varphi(x)
\varphi(y) \hspace{-0.1cm} \rsta{0} + \lsta{0} \hspace{-0.1cm}
\varphi(y) \varphi(x) \hspace{-0.1cm} \rsta{0}, \\
i \Delta_F(x-y) & = & \Theta(x^0-y^0) \lsta{0} \hspace{-0.1cm}
\varphi(x), \varphi(y) \hspace{-0.1cm}
\rsta{0} + \Theta(y^0-x^0) \lsta{0} \hspace{-0.1cm} \varphi(y)
\varphi(x) \hspace{-0.1cm} \rsta{0}, \end{eqnarray}
where $\Theta$ denotes the Heaviside step function,
one obtains immediately from the equations (\ref{e}) and (\ref{f})
induced path-integral representations for \linebreak[3] $
\Delta_+(x-y) \equiv \lsta{0} \hspace{-0.1cm}
\varphi(x) \varphi(y) \hspace{-0.1cm} \rsta{0}$ and $\Delta_-(x-y)
\equiv \lsta{0} \hspace{-0.1cm} \varphi(y) \varphi(x)
\hspace{-0.1cm} \rsta{0}$. This yields an induced path-integral
representation for  $\Delta$
\[ i \Delta(x-y) = \epsilon(y^0-x^0) \left[ \int_0^{\infty}
dN(\tau''-\tau') - \int_{-\infty}^0 dN(\tau''-\tau') \right] \int
{\cal D}p {\cal D}x \exp \left( i \int_{\tau'}^{\tau''} d \tau (p
\dot{x} -N {\cal H}_0) \right), \]
where the function $\epsilon$ is given by eq.~(\ref{g}).
Below more intrinsic path-integral representations will be
derived. \\ \\
Upon canonical quantization, one introduces abstract momentum
eigenstates $\rsta{p}$ by
\begin{equation} \hat{p}_{\mu} \rsta{p} = p_{\mu} \rsta{p},
\end{equation}
where the constraint eq.~(\ref{a}) is imposed as a condition on
the states
\begin{equation} (p^2 -m^2 ) \rsta{p} =0. \end{equation}
The momentum eigenstates can therefore be labeled by the
three-momentum {\bf p} and the sign of $p_0$
\begin{eqnarray} \hat{p}_i \rsta{{\bf p} \pm} & = & p_i \rsta{{\bf
p} \pm}, \mbox{ for } i \in \{1,2,3\}, \\ \hat{p}_0 \rsta{{\bf p}
\pm} & = & \pm ({\bf p}^2 +m^2)^{\frac{1}{2}} \rsta{{\bf p} \pm}.
\end{eqnarray}
Physical states are obtained by a superposition of momentum
eigenstates. \\
By Fouriertransformation one defines Lorentz-invariant
configuration states \cite{Ha92}
\begin{equation} \rsta{x} \equiv \rsta{x +} + \rsta{x -},
\end{equation}
where \begin{eqnarray} \rsta{x +} & \equiv & \frac{1}{(2
\pi)^{\frac{3}{2}}}
\int_{p_0 = \omega_{{\bf \small p}}} \frac{d^3 {\bf
p}}{2 \omega_{\bf \small p}} e^{ipx} \rsta{{\bf p} +}, \\
\rsta{x -} & \equiv & \frac{1}{(2 \pi)^{\frac{3}{2}}} \int_{p_0 =
- \omega_{{\bf \small p}}} \frac{d^3 {\bf p}}{2 \omega_{\bf \small
p}} e^{ipx} \rsta{{\bf p} -}.
\end{eqnarray}
The states of positive and negative energy decouple and may be
considered separately. For the abstract momentum eigenstates with
positive energy $\{ \rsta{{\bf p} +} \} $ an inner product may be
{\em defined} by \begin{equation} \label{i}
\lsta{{\bf p} +} \hspace{-0.05cm}{\bf p'} + \rangle \equiv 2
\omega_{\bf \small p} \delta({\bf p} - {\bf p'}). \end{equation}
Eq.~(\ref{i}) implies the completeness relation
\begin{eqnarray} 1 & = & \int \frac{d^3 {\bf p}}{2 \omega_{\bf
\small p}} \rsta{{\bf p} +} \lsta{{\bf p} +} \\ \label{k} & = & i
\int_{x^0 = const.} d^3 {\bf x} \rsta{x +}
\stackrel{\leftrightarrow}{\partial_0} \lsta{x +}. \end{eqnarray}
To include the states with negative energies the most natural
choice seems to be
\begin{eqnarray} \label{j}
\lsta{{\bf p} i} \hspace{-0.1cm} {\bf p'} j \rangle & \equiv & 2
\omega_{\bf \small p} \delta_{ij} \delta({\bf p} - {\bf p'}),
\mbox{ where } i,j \in \{+,-\}, \\
1 & = & \int \frac{d^3 {\bf p}}{2 \omega_{\bf \small p}} \left[
\rsta{{\bf p} +} \lsta{{\bf p} +} + \rsta{{\bf p} -} \lsta{{\bf p}
-} \right], \nonumber \end{eqnarray}
however as it is shown in \cite{Ha92} eq.~(\ref{k}) is replaced
by the unusual formula
\[ 1 = i \int_{x^0 = const.} d^3 {\bf x} \left[
\rsta{x +} \stackrel{\leftrightarrow}{\partial_0} \lsta{x +} -
\rsta{x -} \stackrel{\leftrightarrow}{\partial_0} \lsta{x -}
\right]. \]
If one wants to stick to the usual relativistic completeness
relation  \[ 1 = i \int_{x^0 = const.} d^3 {\bf x} \left[ \rsta{x
+} \stackrel{\leftrightarrow}{\partial_0} \lsta{x +} + \rsta{x -}
\stackrel{\leftrightarrow}{\partial_0} \lsta{x -} \right] \] one
has to abandon the positive definiteness of the inner product
eq.~(\ref{i}) and instead has to set
\begin{equation} \label{n} \lsta{{\bf p} i} \hspace{-0.1cm}
{\bf p'} j \rangle
\equiv  2 \omega_{\bf \small p} \delta_{i+} \delta_{j+}
\delta({\bf p} - {\bf p'}) - 2 \omega_{\bf \small p} \delta_{i-}
\delta_{j-} \delta({\bf p} - {\bf p'}), \mbox{ where } i,j \in
\{+,-\}. \end{equation}
Now once one has abandoned the
positive definiteness of the inner product eq.~(\ref{i}), there
is no reason to keep $\lsta{{\bf p} \pm} \hspace{-0.1cm}
{\bf p'} \pm \rangle$ real.
Indeed, once the absolute values of $\lsta{{\bf p} \pm}
\hspace{-0.1cm} {\bf p'} \pm \rangle$
are fixed, there remains a phase factor free to
choose
\begin{equation} \label{m} \lsta{{\bf p} i} \hspace{-0.1cm}
{\bf p'} j \rangle
\equiv 2 \omega_{\bf \small p} \delta_{i+} \delta_{j+} \delta({\bf
p} - {\bf p'}) + 2 e^{i \alpha} \omega_{\bf \small p} \delta_{i-}
\delta_{j-} \delta({\bf p} - {\bf p'}), \mbox{ where } i,j \in
\{+,-\}. \end{equation}
Setting $\alpha = 0$ (resp.~$\alpha= \pi$) eq.~(\ref{j})
(resp.~eq.~(\ref{n}))
is readily obtained. The following completeness relations are
straightforward
\begin{eqnarray*} 1 & = & \int \frac{d^3 {\bf p}}{2 \omega_{\bf
\small p}} \left[ \rsta{{\bf p} +} \lsta{{\bf p} +} + e^{- i
\alpha} \rsta{{\bf p} -} \lsta{{\bf p} -}\right], \\
& = & i \int_{x^0 = const.} d^3 {\bf x} \left[ \rsta{x +}
\stackrel{\leftrightarrow}{\partial_0} \lsta{x +} - e^{-i \alpha}
\rsta{x -} \stackrel{\leftrightarrow}{\partial_0} \lsta{x -}
\right]. \end{eqnarray*}
The relativistic propagator turns out to be
\begin{eqnarray} \label{u} {\cal G}_{\alpha}(x'-x) \equiv
\lsta{x'} \hspace{-0.1cm} x \rangle & = & \frac{1}{2}
(\Delta_1(x'-x) + i \Delta(x'-x))
+ \frac{e^{i \alpha}}{2} (\Delta_1(x'-x) - i \Delta(x'-x)) \\
& = & \label{v} \lsta{0} \hspace{-0.1cm} \varphi(x') \varphi(x)
\hspace{-0.1cm} \rsta{0} + e^{i \alpha} \lsta{0} \hspace{-0.1cm}
\varphi(x) \varphi(x') \hspace{-0.1cm} \rsta{0}. \end{eqnarray}
One findes ${\cal G}_0 = \Delta_1$ and ${\cal G}_{\pi} = i
\Delta$.
${\cal G}_{\alpha}$ is a solution of the Klein-Gordon equation for
every $\alpha \in [0, 2 \pi [$.
\section{Path-Integral representation for ${\cal G}_{\alpha}$}
At the end of the last section it was shown that for special
values of $\alpha$ the Green function ${\cal G}_{\alpha}$ is equal
to the Hadamard Green function resp.~to the causal Green
function.
As remarked above $\Delta_1$ possesses a path-integral
representation
eq.~(\ref{f}). Therefore the suggestion in \cite{Ha92} that
possible there
is no sum-over-histories representation for $\Delta$ seems to be
rather unnatural. On the contrary one would expect that after
introducing a suitable phase factor in eq.~(\ref{f}) a
path-integral
representation for ${\cal G}_{\alpha}$ could straightforwardly be
obtained.
The aim of this section is to show that this is indeed the case.
First the result is stated and then checked. \\
The path-integral representation for ${\cal G}_{\alpha}$ is
\begin{eqnarray} \label{z} {\cal G}_{\alpha}(x''-x') = \int_{-
\infty}^{\infty} dN(\tau''-\tau') \int {\cal D}p & \hspace{-0.2cm}
{\cal D}x \hspace{-0.2cm} &
\left[ \left( \prod_{\tau'}^{\tau''} \Theta(p_0) \right) \exp
\left( - i \int_{\tau'}^{\tau''} d \tau (p \dot{x} -N {\cal H}_0)
\right) \right. \\
& & \nonumber \left. + e^{i \alpha} \left( \prod_{\tau'}^{\tau''}
\Theta(-p_0) \right) \exp \left( -i \int_{\tau'}^{\tau''} d \tau
(p \dot{x} -N {\cal H}_0) \right) \right]. \end{eqnarray}
The meaning of the infinite product will become clearer in
eq.~(\ref{r}) below. Setting $\alpha = \pi$ one obtains the
following
path-integral representations for the causal Green function
\begin{eqnarray} \Delta(x'' - x') & = & -i \int_{-\infty}^{\infty}
dN(\tau''-\tau') \nonumber \int {\cal D}p {\cal D}x \left(
\prod_{\tau'}^{\tau''} \epsilon(p_0) \right) \exp \left( -i
\int_{\tau'}^{\tau''} d \tau (p \dot{x} -N {\cal H}_0) \right) \\
& = & -i \int_{-\infty}^{\infty} dN(\tau''-\tau') \int {\cal D}p
{\cal D}x \left( \prod_{\tau'}^{\tau''} \epsilon(p_0) \exp \left(
-i p \dot{x} +iN {\cal H}_0 \right) \right).
\end{eqnarray}
Setting $\alpha =0$ yields again eq.~(\ref{f}).
To prove eq.~(\ref{z}) one uses the well-known constructive
definition of path-integrals by a time slicing procedure. The
measure at each time slice is taken to be the Liouville measure.
Suppressing the integration over
$N(\tau''-\tau')$, the first term in eq.~(\ref{z}) gives
with $\delta = \frac{\tau'' - \tau'}{n+1}$
\begin{eqnarray} \lefteqn{\int {\cal D}p {\cal D}x \left(
\prod_{\tau'}^{\tau''} \Theta(p^0) \right) \exp \left( -i
\int_{\tau'}^{\tau''} d \tau (p \dot{x} -N {\cal H}_0) \right)}
\nonumber \\ \label{r} & & \equiv \lim_{n \to \infty} \int
\prod_{k=1}^n \frac{d^4 x_k d^4 p_k}{(2 \pi)^4} \frac{d^4 p_0}{(2
\pi)^4}
\left( \prod_{h=0}^n \Theta(p^0_h) \exp \left( -ip_h (x_{h+1}-x_h)
+ i \delta N (p_h^2 -m^2) \right) \right). \end{eqnarray}
In eq.~(\ref{r}) the $x_k$'s and $p_k$'s are thought to define a
skeletonized path
in phase space. The $x_k$'s are viewed as the values of
$x(\tau)$ at parameter-time $\tau_k = \tau' + k \delta$, whereas
the $p_k$'s are thought of as the values of $p(\tau)$ at
parameter-time $\frac{1}{2} (\tau_{k+1}+\tau_k) = \tau'
+(k+\frac{1}{2}) \delta$. \\
Setting $n=0$ in (\ref{r}) one obtains a short-time propagator.
It is well known that the short-time evolution of any positive
energy solution of the Klein-Gordon equation is given by this
short-time propagator. The reverse is also true. Namely, given any
wave function, whose short-time evolution is determined by this
short-time propagator, one can show that under reasonable
circumstances this wave function is a solution of
the positive square root of the Klein-Gordon equation. The details
are given in the appendix. \\
Performing the $x_i$-integrations in (\ref{r}) yields
$\delta$-functions  of the
form $\delta(p_i - p_{i-1})$. Therefore all but one
of the \linebreak
           $p$-integrations may be performed to yield
\begin{eqnarray*} \lefteqn{\int {\cal D}p {\cal D}x
\left( \prod_{\tau'}^{\tau''} \Theta(p^0) \right)
\exp \left( -i \int_{\tau'}^{\tau''} d \tau (p \dot{x} -N
{\cal H}_0) \right)}
\\ &=& \int \frac{d^4 p}{(2 \pi)^4} \Theta(p^0) \exp
\left( -i p (x''-x') + i N(\tau''-\tau')(p^2-m^2) \right).
\end{eqnarray*}
The second term of eq.~(\ref{z}) may be manipulated in the same
way. Recalling
\[ (2 \pi) \delta(p^2-m^2) = \int_{-\infty}^{\infty}
dT e^{iT(p^2-m^2)}, \]
and setting $T \equiv N(\tau'' -\tau')$,
comparison with eqs. (\ref{u}), (\ref{v}) and (\ref{x}) -
(\ref{y})
proves eq.~(\ref{z}). \footnote{By the way notice that
$2 \betrag{T} = 2 \betrag{N}(\tau''-\tau')$ is identical with the
so-called `fifth
parameter', which was first introduced by Fock. Feynman used the
fifth parameter to bring the Klein-Gordon-equation in a form
analogous to the Schr\"odinger-equation. Furthermore Feynman stated
a representation of the Feynman propagator $\Delta_F$ as an
integral over the fifth parameter, which can also be obtained
readily from eq.~(\ref{e}) performing the ${\cal D}p{\cal D}x$
integrations \cite{Fe50}.}  \\ \\
The appearance of the Heaviside functions in the path-integral
(\ref{z}) prevents the $p$-integrations from being
straightforwardly
performed. However a little trick makes it possible to perform the
$p$-integrations without affecting the $x$-integrations.
Inserting the identity
\begin{equation} \label{p} \Theta(x) = \lim_{\kappa \to 0}
\frac{1}{2 \pi i} \int_{-\infty}^{\infty} \frac{d \omega}{\omega -
i \kappa} e^{i\omega x} \end{equation}
into the first term in eq.~(\ref{z}) gives
(again suppressing the integration over $N(\tau''-\tau')$),
\begin{eqnarray} \lefteqn{\int {\cal D}p {\cal D}x
\left( \prod_{\tau'}^{\tau''} \Theta(p^0) \right) \exp \left( -i
\int_{\tau'}^{\tau''} d \tau (p \dot{x} -N {\cal H}_0) \right)}
\nonumber \\ & & = \lim_{n \to \infty} \int \prod_{k=1}^n
\frac{d^4 x_k d^4 p_k}{(2 \pi)^4} \frac{d^4 p_0}{(2 \pi)^4}
\left( \prod_{h=0}^n \Theta(p^0_h) \exp \left( -ip_h (x_{h+1}-x_h)
+ i \delta N (p_h^2 -m^2) \right) \right) \\
& & \nonumber = \lim_{n \to \infty} \int \prod_{k=1}^n \frac{d^4
x_k d^4 p_k}{(2 \pi)^4} \frac{d^4 p_0}{(2 \pi)^4}
\left[ \prod_{h=0}^n \lim_{\kappa_h \to 0} \frac{d \omega_h}{2 \pi
i(\omega_h -i \kappa_h)} \exp \left( ip_h^0 \omega_h - i p_h
(x_{h+1}-x_h) + i \delta N (p_h^2 -m^2) \right) \right].
\end{eqnarray}
After interchanging the integration over $p_h$ and the limit
$\kappa_h \to 0$, the Gaussian integral over $p_h$ is
straightforward. Performing all $p$-integrations gives the result
\footnote{To make the integrations well defined, one has to add to
$N$ a suitable imaginary part: for the $p_0$-integration: $N
\mapsto N +i s^2$ and for the $p_j$-integration ($j =1,2,3$): $N
\mapsto N -i s^2$. After the integration $s^2$ is send to zero.}
\begin{eqnarray} \lefteqn{\int {\cal D}p {\cal D}x \left(
\prod_{\tau'}^{\tau''} \Theta(p^0) \right) \exp \left( -i
\int_{\tau'}^{\tau''} d \tau (p \dot{x} -N {\cal H}_0) \right)}
\nonumber \\ & & = \frac{1}{(2 \pi)^4} \lim_{n \to \infty} \int
\prod_{k=1}^n \frac{d^4 x_k}{(2 \pi)^4} \prod_{h=0}^n \left[
\lim_{\kappa_h \to 0} \frac{d \omega_h}{2 \pi i(\omega_h -i
\kappa_h)} i \epsilon(N) \left(\frac{\pi}{\delta N} \right)^2
\right. \nonumber \\ & & \hspace*{0.4cm} \nonumber \left. \times
\exp \left( \frac{-i}{4 \delta N}
\left[ (x_{h+1}-x_h)^2 -2 (x^0_{h+1}-x^0_h) \omega_h +
\omega_h^2 + 4 \delta^2 N^2 m^2 \right] \right) \right],
\end{eqnarray}
where $\epsilon$ was defined in (\ref{g}).
Application of Cauchy's integral formula and taking $\kappa_h \to
0$ yields
\begin{eqnarray} \lefteqn{\int {\cal D}p {\cal D}x \left(
\prod_{\tau'}^{\tau''} \Theta(p^0) \right) \exp \left( -i
\int_{\tau'}^{\tau''} d \tau (p \dot{x} -N {\cal H}_0) \right)}
\nonumber \\ & & = \frac{1}{(2 \pi)^4} \lim_{n \to
\infty} \int \prod_{k=1}^n \frac{d^4 x_k}{(2 \pi)^4} \prod_{h=0}^n
\left[ i \epsilon(N) \left(\frac{\pi}{\delta N} \right)^2 \Theta
\left( \frac{x_{h+1}^0-x^0_h}{\delta N} \right) \exp \left( -i
\frac{(x_{h+1}-x_h)^2}{4 \delta N}  - i \delta N m^2
\right) \right] \nonumber \\
\label{q} & & \equiv \int \tilde{\cal D} x \hspace{0.25em} \Theta
\left(
\frac{\dot{x}_0}{N} \right) \exp \left( -i \int_{\tau'}^{\tau''}
d \tau \left( \frac{\dot{x}^2}{4N} + N m^2 \right) \right),
\end{eqnarray}
where the symbolic measure $\tilde{\cal D} x$ is defined as
follows
\begin{equation} \label{sm} \tilde{\cal D} x = \frac{1}{(2 \pi)^4}
\lim_{n
\to \infty} \left( \prod_{k=1}^n \frac{d^4 x_k}{(2 \pi)^4} \right)
\left(\frac{\pi}{\delta N} \right)^{2n+2} (i \epsilon(N))^{n+1}.
\end{equation}
For notational simplicity we suppress the (formal) infinite
product over all times in front of the Heaviside function in
eq.~(\ref{q}).
Remarkably enough after performing the $p_h$-integration the
factor $\Theta(p^0_h)$ turns into a factor $\Theta(
\frac{x_{h+1}^0-x_h^0}{\delta N})$. Thus if $N>0$
a particle with positive energy is traveling  forward in time and
if $N<0$ a particle with positive energy is traveling  backward
in time. \\
For the second term in eq.~(\ref{z}) the $p$-integrations may be
performed in the same way as for the first term. The factor
$\Theta(-p_h^0)$ turns into $\Theta( -\frac{x_{h+1}^0-
x_h^0}{\delta N})$. Therefore a particle with negative
energy is traveling  forward (resp.
backward) in time, if $N<0$ (resp. $N>0$). Therefore all
trajectories of physical interest are included if one considers
the case $N>0$ only. These results have been obtained already in
\cite{Te82} by completely different means considering the
classical equations of motion of the relativistic particle. \\
The final result for the path-integral representation of ${\cal
G}_{\alpha}$ in configuration space is
\begin{equation} \label{ghjk} {\cal G}_{\alpha}(x'' -x') = \int_{-
\infty}^{\infty} dN(\tau''-\tau') \left\{ \int \tilde{\cal D} x
\left[ \Theta \left( \frac{\dot{x}_0}{N} \right) + e^{i \alpha}
\Theta \left( \frac{-\dot{x}_0}{N} \right) \right] \exp \left( -i
\int_{\tau'}^{\tau''} d \tau \left( \frac{\dot{x}^2}{4N} + N m^2
\right) \right) \right\}. \end{equation}
Especially for $\Delta$ and $\Delta_1$ one obtains
\begin{eqnarray}
i \Delta(x''-x') & = & \label{zb} \int_{-\infty}^{\infty}
dN(\tau''-\tau')
\int \tilde{\cal D} x \hspace{0.25em} \epsilon \left(
\frac{\dot{x}_0}{N} \right) \exp \left( -i \int_{\tau'}^{\tau''} d
\tau \left( \frac{\dot{x}^2}{4N} + N m^2 \right) \right), \\
\label{za} \Delta_1(x''-x') & = & \int_{-\infty}^{\infty}
dN(\tau''-\tau') \int
\tilde{\cal D}x \exp \left( -i \int_{\tau'}^{\tau''} d \tau \left(
\frac{\dot{x}^2}{4N} + N m^2 \right) \right).
\end{eqnarray}
\subsubsection*{BRST Path integral representation for ${\cal
G}_{\alpha}$}
There is another path integral representation in phase space for
the Green functions ${\cal G}_{\alpha}$, which uses techniques
from BRST theory. An introduction to BRST symmetry as needed here
can be found in \cite{He92}. \\
The idea behind what follows is to derive a path integral
representation for the first and second term in eq.~(\ref{v})
separately. Eq. (\ref{z}) induces a path integral representation
for $\Delta_+(x'-x) = \lsta{0} \hspace{-0.1cm} \varphi(x')
\varphi(x) \hspace{-0.1cm} \rsta{0}$, which can also written as
\begin{equation}
\Delta_+(x''-x') = \int_{-\infty}^{\infty} dN(\tau''-\tau')
\int {\cal D}p {\cal D}x
\left( \prod_{\tau'}^{\tau''} \Theta(-p_0) \right) \exp
\left( i \int_{\tau'}^{\tau''} d \tau (p \dot{x} -N {\cal H}_0)
\right). \end{equation}
The infinite product of Heaviside-functions indicate that in BRST
theory one has to obey both the constraint eq.~(\ref{a}) and the
further constraint $p_0(\tau) < 0$, for all $\tau$.
The simplest possibility to incorporate both constraints into a
single constraint function is to take the `square root' of the
constraint eq.~(\ref{a}):
\begin{equation} \label{t} \tilde{{\cal H}}_0 \equiv p_0^2 + p_0
\omega_{\bf \small p} =0, \end{equation}
where $\omega_{\bf \small p}$ is defined by eq.~(\ref{om}).
It should be stressed that it is in general not allowed to choose
an
equivalent constraint, e.g. $p_0 + \omega_{\bf \small p} =0$,
because the resulting BRST path integral does in general not
coincidate with $\Delta_+$ and furthermore it does not even have
the same dimension as $\Delta_+$. This feature is not well
understood and the author hopes to come back to it in a future
publication. \\
With the constraint (\ref{t}) at hand the derivation of the BRST
path integral is straightforward. Therefore only the bare
essentials are given here. The complete story can be found in
\cite{He92}. \\
In a first step one introduces a ghost pair $(\eta, {\cal P})$
associated with the constraint $\tilde{\cal H}_0 =0$. $\eta$ and
$\cal P$ are canonically conjugate anticommuting variables. $\cal
P$ is also called the ghost momentum. The Lagrange multiplier $N$
introduced to incorporate the constraint into the action (see
eq.~(\ref{b})) is viewed as a dynamical variable. As a further
constraint the momentum $\Pi$ conjugate to $N$ is required to
vanish on the constraint surface
\[ \Pi = 0. \]
Now a second anticommuting ghost pair $(\rho, \bar{C})$ - called
antighosts - associated with the constraint $\Pi =0$ has to be
introduced. \\
Finally the extended phase space is defined to be the space
spanned by the original phase space coordinates $p_{\mu}$ and
$x^{\mu}$ together with the ghost variables $\eta, {\cal P}, \rho,
\bar{C}$ and with the multiplier $N$ and its conjugate momentum
$\Pi$. The Poisson-bracket is extended as follows
\[ \{ {\cal P}, \eta \} = \{ \rho, \bar{C} \} = \{\Pi, N \} = -1,
\]
and all other Poisson-brackets involving the new variables $\eta,
{\cal P}, \rho, \bar{C}, N, \Pi$ are taken to vanish.
The extended action appropiate to the dynamics in extended phase
space is called the gauge-fixed action and is given by
\begin{equation} \label{s} S_E(p_{\mu}, x^{\mu}, \eta, {\cal P},
\rho, \bar{C}, N, \Pi) = \int_{\tau'}^{\tau''} d \tau ( p_{\mu}
\dot{x}^{\mu} + \dot{\eta} {\cal P} + \dot{N} \Pi + \dot{\bar{C}}
\rho - \{K, \Omega \}). \end{equation}
Here as usual $\Omega$ denotes the BRST charge
\[ \Omega = \eta \tilde{\cal H}_0 - i \rho \Pi. \]
The action (\ref{s}) is invariant under transformations generated
by the BRST charge $\Omega$. The factor $-i$ is convention.
The BRST charge $\Omega$ is nilpotent $\{ \Omega, \Omega \} = 0$.
$K$ is called the gauge-fixing fermion. \\
Now the BRST path integral is the formal expression
\begin{equation} \label{BR} PI = \int {\cal D} p {\cal D} x {\cal
D} \eta
{\cal D P D} N {\cal D} \Pi {\cal D} \rho {\cal D} \bar{C} \exp(i
S_E). \end{equation}
This path integral is thought to be defined by a time slicing
procedure with the boundary conditions eq.~(\ref{w}) together with
\[ \Pi = \bar{C} = \eta = 0 \]
at the boundaries. The measure at each time slice is taken to be
the Liouville measure. The famous Fradkin-Vilkovisky-Theorem
\cite{He92,Fr75} states that we are free to
choose $K$ as \[ K = - {\cal P} N, \] without changing the value of
the expression in (\ref{BR}).
In \cite{He92} it is shown that this choice is compatible with
eq.~(\ref{d}), that is, the equations of motion derived from the
action (\ref{s}) imply $\dot{N}=0$. \\
With the choice $K=-{\cal P}N$ the
integrations over the ghosts and antighosts decouple from the
other integrations and may be performed separately
\begin{eqnarray} \label{gh} \int {\cal D}\eta {\cal D P D} \rho
{\cal D}
\bar{C} \exp(iS_E) & \equiv & \lim_{n \to \infty} \int
\prod_{h=1}^n d \eta_h \int \prod_{h=1}^n d \bar{C}_h
\int \prod_{h=0}^n d \rho_h \int \prod^n_{h=0} d {\cal P}_h \\
& & \hspace{0.2cm} \times \exp \left[ \sum_{h=0}^n
\left( i(\eta_{h+1}-\eta_h) {\cal P}_h
+ i(\bar{C}_{h+1} - \bar{C}_h) \rho_h + \delta {\cal P}_h \rho_h
\right) \right]. \nonumber
\end{eqnarray}
Strictly speaking the left hand side of eq.~(\ref{gh}) is
defined through the right hand side.
Expanding the exponential function and bearing in mind the
definition of integration of anticommuting numbers, that is $\int
d \kappa \kappa \equiv 1$ and $\int d \kappa  \equiv 0$, one
easily sees that only the terms containing the ${\cal P}_h,
\rho_h$ in the form $\prod_h {\cal P}_h \rho_h$ survive the ${\cal
P}, \rho$-integrations. Properly adjusting all prefactors after
performing the ${\cal P}$ and $\rho$-integrations yields
\begin{eqnarray} & (-1)^{\frac{n^2+n}{2}}
& \left[ \delta^{n+1} - \delta^n \sum_{h=0}^n
(\bar{C}_{h+1}-\bar{C}_h) (\eta_{h+1}-\eta_h) \right. \nonumber \\
& & \label{hg} \hspace*{0.2cm} + \delta^{n-1} \sum_{h,j=0 \atop
j < h}^n(\bar{C}_{h+1}-
\bar{C}_h)(\eta_{h+1}-\eta_h)(\bar{C}_{j+1}-\bar{C}_j)(\eta_{j+1}-
\eta_j) \pm ... + \\ \nonumber & & \hspace*{0.3cm} + \delta (-1)^n
\sum_{h_i=0 \atop h_1<h_2< \cdot \cdot \cdot <h_n}^n
(\bar{C}_{h_1+1}-
\bar{C}_{h_1})(\eta_{h_1+1}-\eta_{h_1}) \cdot \cdot \cdot
(\bar{C}_{h_n+1}-\bar{C}_{h_n})(\eta_{h_n+1}-\eta_{h_n}) \Bigg].
\end{eqnarray}
The factor $(-1)^{\frac{n^2+n}{2}}$ results
from the various commutations of the differentials $d\rho$ and $d
\cal P$.
Only the last term in eq.~(\ref{hg}) contribute to the $\eta$ and
$\bar{C}$-integrations. An easy combinatorial argument yields
that eq.~(\ref{gh}) equals \begin{equation} \label{re}
(-1)^n (-1)^{\frac{n^2+n}{2}}\delta (n+1) (-1)^{\frac{n^2-
n}{2}} = (n+1) \delta = \tau'' - \tau'.
\end{equation}
(Pedestrians may check this by induction.)
The $\Pi,N$-integrations in eq.~(\ref{BR}) are quite easy
\begin{eqnarray}
\int {\cal D}N {\cal D} \Pi \exp \left( i \int d \tau \dot{N} \Pi
\right) & = &
\int \prod_{h=1}^n \frac{dN_h d \Pi_h}{2 \pi} dN_0 \exp \left( i
\sum_{h=0}^n (N_{h+1} -N_h)\Pi_h \right) \nonumber \\
& = & \int \prod_{h=0}^n dN_h \delta \left( N_{h+1} -N_h \right).
\nonumber \end{eqnarray}
So one finally arrives at
\begin{equation} \label{pi}
PI = \int_{-\infty}^{\infty} dN (\tau'' - \tau') \int {\cal D} p
{\cal D} x \exp \left( i \int_{\tau'}^{\tau''} d \tau (p_{\mu}
\dot{x}^{\mu} - N \tilde{\cal H}_0) \right). \end{equation}
An analysis similar to the above has already been carried out in
\cite{Ha88} for the relativistic particle with the constraint
${\cal H}_0$.
Now it is straightforward to check that $PI = 2 \Delta_+$:
Setting $T \equiv N(\tau''-\tau')$ one obtains
\begin{eqnarray}
PI & = & \int_{-\infty}^{\infty} dN(\tau''-\tau') \int
\prod_{h=1}^n \frac{d^4p_h d^4x_h}{(2 \pi)^4}
\frac{d^4p_0}{(2 \pi)^4} \exp \left\{ i
\sum_{h=0}^n \left[ p_h(x_{h+1}-x_h) - \delta N p_h^0 (p_h^0 +
\omega_{\bf \small p_{\rm h}}) \right] \right\} \nonumber \\
& = & \int_{-\infty}^{\infty} dT \int \frac{d^4p}{(2 \pi)^4}
\exp \left\{ -i p(x''-x') - i T (p^0p^0 - p^0
\omega_{\bf \small p}) \right\} \nonumber \\
& = & \int \frac{d^4p}{(2 \pi)^3} \delta(p^0p^0-p^0 \omega_{\bf
\small p}) \exp(-i p(x''-x')) \nonumber \\
&=& \int_{p^0 = + \omega_{\bf \small p}} \frac{d^3 {\bf p}}{(2
\pi)^3 \omega_{\bf \small p}} \exp(-ip(x''-x')) \\
&=& 2 \Delta_+(x''-x').
\end{eqnarray}
The final result of this subsection is therefore
\begin{eqnarray}
\Delta_+(x''-x') & = & \frac{1}{2} \int_{-
\infty}^{\infty} dN(\tau''-\tau') \int {\cal D}p {\cal D}x
 \exp \left( i \int_{\tau'}^{\tau''} d \tau (p \dot{x} -N
\tilde{\cal H}_0) \right) \\
& = & \frac{1}{2} \int {\cal D} p {\cal D} x {\cal D} \eta
{\cal D P D} N
{\cal D} \Pi {\cal D} \rho {\cal D} \bar{C} \exp(i S_E).
\end{eqnarray}
A path integral representation for $\Delta_-$ can be obtained
similar using the constraint $\tilde{\cal H}_0^- \equiv p_0^2-p_0
\omega_{\bf \small p}$.
Putting the path integral representations for $\Delta_+$ and
$\Delta_-$ together yields a path-integral representation for
${\cal G}_{\alpha}$. This concludes the derivation of our new
BRST-path-integral representation for ${\cal G}_{\alpha}$.
\section{Composition laws for ${\cal G}_{\alpha}$}
The main goal in \cite{Ha92}was to derive the relativistic
composition laws for the considered Green functions from their
path integral representations. For this task the path
decomposition expansion (PDX) was used \cite{Au85}. \\
In this section the same thing will be done for the Green
functions ${\cal G}_{\alpha}$.
To this end the path decomposition expansion has to be generalized
to be applicable to path-integrals of the form eq.~(\ref{q}).
\subsubsection*{The Path Decomposition Expansion}
The path decomposition expansion allows to express the dynamics in
full configuration space through the dynamics in two disjunct
regions of configuration space separated by a surface $\Sigma$.
The reader not familiar with the PDX is referred to the references
\cite{Ha92,Au85}. \\
In the following the path decomposition expansion for path
integrals of the form
\begin{equation} \label{x2} K \left( x(\tau'), x(\tau''),N,\tau''-
\tau' \right)
\equiv \int \tilde{\cal D} x \hspace{0.25em} \Theta \left(
\epsilon(N) \dot{x}_0 \right) \exp \left( i \int_{\tau'}^{\tau''}
d \tau \left( \frac{\dot{x}^2}{4N} + N m^2 \right) \right)
\end{equation}
will be derived. The measure $\tilde{\cal D} x$ was defined above
in eq.~(\ref{sm}) and the function $\epsilon$ was defined in
eq.~(\ref{g}). \\
The following derivation of the PDX is a modification
of the original derivation given in \cite{Au85}. The PDX is
derived in the Euclidian regime. The Euclidian version of
(\ref{x2}) is obtained by rotating both $\tau$ and $x^0$
in the following manner
\[ \tau \mapsto \tau_E = - \epsilon(N) i \tau; \hspace{1cm} x^0
\mapsto x^0_E = - i x^0. \]
In the following the index `$E$' will be suppressed.
Let $\Sigma(\tilde{x}^0)$ be the surface in configuration space of
constant $x^0$-coordinate $x^0 = \tilde{x}^0$, where $\tilde{x}^0$
lies between $x^0(\tau'')$ and $x^0(\tau')$. \\
First consider the Wick-rotated path integral
\begin{eqnarray} & & \label{oe} \int \tilde{\cal D} x
\hspace{0.25em} \Theta \left( \epsilon(N) \dot{x}_0 \right) \exp
\left( - \epsilon(N) \int_{\tau'}^{\tau''} d
\tau \left( \frac{\sum_{\mu=0}^3 (\dot{x}^{\mu})^2}{4N} + N m^2
\right) \right) \\
& & \nonumber \equiv \lim_{n \to \infty}
\frac{-i}{(2 \pi)^4}
\int \prod_{k=1}^n \frac{d^4 x_k}{(2 \pi)^4} \prod_{h=0}^n
\left[ \left(\frac{\pi}{\delta N} \right)^2 \Theta \left(
\frac{x_{h+1}^0-x^0_h}{\delta} \right) \exp \left( -
\frac{\sum_{\mu=0}^3(x^{\mu}_{h+1}-x^{\mu}_h)^2}{4 \delta
\epsilon(N) N} - \epsilon(N) \delta N m^2
\right) \right], \end{eqnarray}
where as above $\delta = \frac{\tau''-\tau'}{n+1}$.
Now because of the factor $\Theta(\dot{x}^0)$ every path in the
sum (\ref{oe}) crosses $\Sigma(\tilde{x}^0)$ once and only once.
Let ${\cal S}_m$ be the set of paths, which fulfill
$x^0_m \leq \tilde{x}^0 < x^0_{m+1}$. The
contribution of ${\cal S}_m$ to the sum (\ref{oe}) is
\begin{eqnarray} & & \frac{-i}{(2 \pi)^4}
\int \prod_{k=1}^n \frac{d^3 x_k}{(2 \pi)^4}
\int_{x^0(\tau')}^{\tilde{x}^0} dx^0_1 \int_{x^0_1}^{\tilde{x}^0}
dx_2^0 \cdot \cdot \cdot \int_{x^0_{m-1}}^{\tilde{x}^0} dx_m^0
\int_{\tilde{x}^0}^{x^0(\tau'')} dx_{m+1}^0
\int_{x^0_{m+1}}^{x^0(\tau'')} dx_{m+2}^0 \cdot \cdot \cdot
\int_{x^0_{n-1}}^{x^0(\tau'')} dx_n^0 \nonumber \\
 & & \hspace{1.5cm} \times \prod_{h=0}^n
\left[ \left(\frac{\pi}{\delta N} \right)^2 \exp \left( -
\frac{(x^0_{h+1}-x^0_h)^2}{4 \delta \epsilon(N) N} -
\frac{\sum_{i=1}^3(x^i_{h+1}-x^i_h)^2}{4
\delta \epsilon(N) N} - \epsilon(N) \delta N m^2
\right) \right]. \label{yz} \end{eqnarray}
The integral in eq.~(\ref{yz}) is in the sequel denoted by
$K_m^n(x(\tau'), x(\tau''), N, -i \epsilon(N) (\tau''-\tau'))$.
The hole path integral (\ref{oe}) is obtained by summing over all
$m$ and taking $n \to \infty$.
Therefore the expression in eq.~(\ref{oe}) equals $\lim_{n \to
\infty} \sum_m K_m^n$.
Now the following identity is needed \cite{Au85}
\begin{eqnarray} \lefteqn{\left(\frac{1}{4 \delta \pi
\epsilon(N) N} \right)^2 \exp \left( - \frac{\sum_{\mu=0}^3
(x^{\mu}_{m+1}-x^{\mu}_m)^2}{4 \delta \epsilon(N) N} \right)}
\nonumber \\ &
& = \int_0^{\delta} d\kappa \int_{\Sigma(\tilde{x}^0)} d^3{\bf
\tilde{x}}
\left(\frac{1}{4 \kappa \pi \epsilon(N) N} \right)^2 \exp \left( -
\frac{\sum_{\mu=0}^3(\tilde{x}^{\mu}-x^{\mu}_m)^2}{4 \kappa
\epsilon(N) N} \right) 2\epsilon(N) N \nonumber \\
& & \hspace{2.5cm} \times \left(\frac{1}{4 (\delta-\kappa) \pi
\epsilon(N)N} \right)^2 \frac{\partial}{\partial x^0}\exp \left( -
\frac{\sum_{\mu=0}^3(x^{\mu}_{m+1}-x^{\mu})^2}{4 (\delta-\kappa)
\epsilon(N) N} \right) \vline_{x=\tilde{x}}. \label{xy}
\end{eqnarray}
Inserting eq.~(\ref{xy}) into eq.~(\ref{yz}) yields
\begin{eqnarray} \lefteqn{K_m^n \left( x', x'', N, -i
\epsilon(N) (\tau''-\tau') \right) =} \\
\nonumber & & i \int_0^{\delta} d \kappa
\int_{\Sigma(\tilde{x}^0)} d^3{\bf
\tilde{x}} K^m_{m+1} \left( x', \tilde{x}, N, -i \epsilon(N)
(m\delta+\kappa) \right) 2 \betrag{N} \frac{\partial}{\partial
x^0}
K^{n-m}_0 \left( x, x'', N, - i\epsilon(N) (\tau''-\tau'-m \delta-
\kappa) \right)
\vline_{x=\tilde{x}}. \end{eqnarray}
Summing over all $m$, taking the limit $n \to \infty$ and rotating
back to real time finally yields the path decomposition expansion
for $K$
\begin{equation}  K \left( x', x'', N, (\tau''-\tau')
\right) = 2 i N
\int_{\tau'}^{\tau''} d t \int_{\Sigma(\tilde{x}^0)} d^3 {\bf
\tilde{x}} \hspace{0.25em} K \left( x', \tilde{x}, N, t
\right)
\frac{\partial}{\partial x^0} K \left( x, x'', N, (\tau''-
\tau'-t ) \right) \vline_{x=\tilde{x}}.
\end{equation}
The PDX can be written more symmetrically
\begin{equation} \label{PDX}  K \left( x', x'', N,
(\tau''- \tau') \right) = i N \int_{\tau'}^{\tau''} d t
\int_{\Sigma(\tilde{x}^0)} d^3
{\bf \tilde{x}} \hspace{0.25em} \left\{ K \left( x',
x, N, t \right)
\stackrel{\leftrightarrow}{\partial_0} K \left( x, x'', N,
(\tau''-\tau'-t ) \right) \right\} \vline_{x=\tilde{x}}.
\end{equation}
This PDX differs from the usual PDX in that no restricted Green
function appears, which clearly is due to the appearance of the
infinite product of Heaviside functions in the sum-over-histories
eq.~(\ref{x2}). \\
For path integrals of the form
\begin{equation} \label{x3} K' \left( x', x'',N,\tau''-
\tau' \right) \equiv \int \tilde{\cal D} x \hspace{0.25em} \Theta
\left( - \epsilon(N) \dot{x}_0 \right) \exp \left( i
\int_{\tau'}^{\tau''} d
\tau \left( \frac{\dot{x}^2}{4N} + N m^2 \right) \right)
\end{equation}
the PDX has the following form
\begin{equation} \label{PDX-}  K'(x', x'', N,
(\tau''-\tau')) = - i N \int_{\tau'}^{\tau''} d t
\int_{\Sigma(\tilde{x}^0)} d^3 {\bf \tilde{x}} \hspace{0.25em}
\left\{ K' \left( x', x, N, t \right)
\stackrel{\leftrightarrow}{\partial_0} K' \left( x, x'', N,
(\tau''-\tau'-t ) \right) \right\} \vline_{x=\tilde{x}}.
\end{equation}
\subsubsection*{Composition laws}
First the composition law for $\Delta_+$  will be derived. The
reasoning is similar to that in \cite{Ha92}. Setting
$T \equiv N(\tau''-\tau')$ the path integral representation
obtained above can written as
\begin{equation} \Delta_+(x''-x') = \int_{-\infty}^{\infty} dT
\int \tilde{\cal D} x \hspace{0.25em} \Theta \left( -\epsilon(T)
\dot{x}^0 \right) \exp \left( i \int_0^T dt \left(
\frac{\dot{x}^2}{4} + m^2 \right) \right). \end{equation}
Insertion of the PDX eq.~(\ref{PDX-}) yields
\begin{equation} \Delta_+(x''-x') = i \int_{-\infty}^{\infty} dT
\int_0^T d \tilde{t} \int_{\Sigma(\tilde{x}^0)} d^3
{\bf \tilde{x}} \hspace{0.25em} \delta_+(x''-\tilde{x}, T,
\tilde{t}) \stackrel{\leftrightarrow}{\partial_0}
\delta_+(\tilde{x}-x', \tilde{t}, 0),
\end{equation}
where \[ \delta_+(x''-x',T_1,T_0) \equiv \int
\tilde{\cal D} x \hspace{0.25em} \Theta \left( -\epsilon(T_1-T_0)
\dot{x}^0 \right) \exp
\left( i \int_{T_0}^{T_1} dt \left(
\frac{\dot{x}^2}{4} + m^2 \right) \right). \]
The composition law of $\Delta_+$ follows immediately after the
substitution $v \equiv T - \tilde{t}$ and $u \equiv \tilde{t}$.
\begin{eqnarray} \label{Dp} \Delta_+(x''-x') & = &
i \int_{-\infty}^{\infty} dv
\int_{-\infty}^{\infty} du \int_{\Sigma(\tilde{x}^0)} d^3
{\bf \tilde{x}} \hspace{0.25em} \delta_+(x''-\tilde{x}, v, 0)
\stackrel{\leftrightarrow}{\partial_0}
\delta_+(\tilde{x}-x', u, 0) \nonumber
\\ & = & i \int_{\Sigma(\tilde{x}^0)}
d^3 {\bf \tilde{x}} \hspace{0.25em} \Delta_+(x''-\tilde{x})
\stackrel{\leftrightarrow}{\partial_0}
\Delta_+(\tilde{x} - x'). \end{eqnarray}
In the same way eq.~(\ref{PDX}) yields for $\Delta_-$
\begin{equation} \label{Dm} \Delta_-(x''-x') = -i
\int_{\Sigma(\tilde{x}^0)}
d^3 {\bf \tilde{x}} \hspace{0.25em} \Delta_-(x''-\tilde{x})
\stackrel{\leftrightarrow}{\partial_0}
\Delta_-(\tilde{x} - x'). \end{equation}
Using that $\Delta_+$ and $\Delta_-$ are orthogonal
\[ \int_{\Sigma(\tilde{x}^0)} d^3 {\bf \tilde{x}} \hspace{0.25em}
\Delta_{\pm}(x''-
\tilde{x}) \stackrel{\leftrightarrow}{\partial_0}
\Delta_{\mp}(\tilde{x}-x') =0 \]
the 'composition law` for ${\cal G}_{\alpha}$ follows
\begin{equation} \label{cl} \int_{\Sigma(\tilde{x}^0)} d^3 {\bf
\tilde{x}} \hspace{0.25em} {\cal G}_{\alpha}(x''-\tilde{x})
\stackrel{\leftrightarrow}{\partial_0} {\cal
G}_{\alpha}(\tilde{x}-
x') = - i \Delta_+(x''-x') + i e^{2i \alpha} \Delta_-(x''-x'),
\end{equation}
which can also be written as
\begin{equation} {\cal G}_{\alpha}(x''-x') = i
\int_{\Sigma(\tilde{x}^0)} d^3 {\bf \tilde{x}}
\hspace{0.25em} {\cal G}_{\frac{\pi+\alpha}{2}}(x''-\tilde{x})
\stackrel{\leftrightarrow}{\partial_0} {\cal
G}_{\frac{\pi+\alpha}{2}}(\tilde{x}-x').
\end{equation}
The well known composition laws for $\Delta$ and $\Delta_1$ are
obtained from eq.~(\ref{cl}) by setting $\alpha=0$ resp.~$\alpha=\pi$.
\begin{eqnarray}
\Delta(x''-x') & = & -\int_{\Sigma(\tilde{x}^0)} d^3 {\bf \tilde{x}}
\hspace{0.25em} \Delta(x''- \tilde{x})
\stackrel{\leftrightarrow}{\partial_0}
\Delta(\tilde{x}-x'), \\
\Delta(x''-x') & = & +\int_{\Sigma(\tilde{x}^0)} d^3 {\bf
\tilde{x}} \hspace{0.25em} \Delta_1(x''-\tilde{x})
\stackrel{\leftrightarrow}{\partial_0} \Delta_1(\tilde{x}-x').
\end{eqnarray}
The composition laws for other Green functions of free scalar field
theory - e.g.~the  composition law for the Feynman propagator
$\Delta_F$ - follow also from eq.~(\ref{Dp})  and (\ref{Dm}).
\section{Conclusion}
In this paper we have obtained two different new path integral
representations in phase space for the Green functions $\Delta_+$
and $\Delta_-$, respectively. The key point concerning the first
path integral representation (\ref{z})
is that its local measure contains an infinite product of
Heaviside functions. The second new path integral representation
is the BRST path integral in extended phase space associated with
a special choice of the constraint.
The central result of this paper is
the sum-over-histories representation (\ref{q})
for $\Delta_+$ and $\Delta_-$, respectively. \\
Putting the path integral and sum-over-histories representations
for $\Delta_+$ and $\Delta_-$ together we obtained the path
integral (\ref{z}) and the
sum-over-histories representations (\ref{ghjk}) for the
${\cal G}_{\alpha}$'s,
introduced in eqs.~(\ref{u}), (\ref{v}). \\
Setting $\alpha = \pi$ we achieved the main goal of this paper,
namely obtaining the sum-over-histories representation (\ref{zb})
for the causal Green function in compact form.  \\
A rather complicated sum-over-histories representation for the
causal Green function has already
been discussed in \cite{An93}. However we do not expect this
representation to be of great use in practice. \\
In the last section we derived a modified path decomposition expansion.
Finally we derived the composition laws
for $\Delta_{\pm}$ from their sum-over-histories representation
using the modified path decomposition
expansion. \\
Furthermore en passant we improved many results already obtained in
\cite{Ha92}. \\
An extensive discussion of the physical motivations underlying
this paper and of the relevance for quantum cosmology of the
obtained results can be found in \cite{Ha92}.
\subsubsection*{Acknowledgments}
I want to thank C.~Grosche and F.~Steiner for reading the
manuscript and for valuable remarks.
\newpage
\begin{appendix}
\section*{Appendix}
In this appendix we explain how the Klein-Gordon equation is
reobtained from (\ref{r}). \\
As is well known, the time evolution of a solution $\varphi^+$
of the Klein-Gordon equation with positive energy is given by the
formula
\[ \varphi^+(x_2) = i \int d^3 {\bf x} \Delta_+(x_2-x_1)
\stackrel{\leftrightarrow}{\partial_0} \varphi^+(x_1). \]
Now let us forget for a moment this result and investigate the
short-time propagator obtained from (\ref{r}) by setting $n=0$
\[ \int dT \int \frac{d^4 p}{(2 \pi)^4}
\left( \Theta(p^0) \exp \left( -ip (x''-x')
+ i T (p^2 -m^2) \right) \right). \]
Imagine a wave function, whose short-time evolution is given by
\[ \varphi({\bf x}', x_0 + \epsilon) = \int dT \int \frac{d^4
p}{(2 \pi)^4} \int d^3 {\bf x} \left( \Theta(p^0) \exp \left(
-i p^0 \epsilon + i {\bf p} ({\bf x}'-{\bf x})
+ i T (p^2 -m^2) \right) P(\stackrel{\leftarrow}{\partial_0},
\stackrel{\leftarrow}{\partial_k},
\stackrel{\rightarrow}{\partial_0},
\stackrel{\rightarrow}{\partial_k}) \varphi({\bf x}, x_0)
\right), \]
where $\epsilon >0$ and $ P(\stackrel{\leftarrow}{\partial_0},
\stackrel{\leftarrow}{\partial_k},
\stackrel{\rightarrow}{\partial_0},
\stackrel{\rightarrow}{\partial_k}) $ is a polynomial in the
derivative operators $\stackrel{\leftarrow}{\partial_{\mu}}$ and
$\stackrel{\rightarrow}{\partial_{\mu}}$.
Expanding the left and the right hand side in a power series in
$\epsilon$ and equating equal powers in $\epsilon$ one obtains
\begin{eqnarray*}
\frac{\partial^n \varphi}{\partial x_0^n}({\bf x}', x_0) & = & \int
dT \int \frac{d^4 p}{(2 \pi)^4} \int d^3 {\bf x} \left(
\Theta(p^0) e^{-i p^0 \epsilon + i {\bf p} ({\bf x}'-{\bf x})
+ i T (p^2 -m^2) } P(i p^0, -i p_k,
\stackrel{\rightarrow}{\partial_0},
\stackrel{\rightarrow}{\partial_k}) \varphi({\bf x}, x_0)
\right), \\ & = & \int d^3 {\bf x} \frac{d^4 p}{(2 \pi)^3}
\frac{\delta(p^0 - \omega_{\bf p})}{2 \omega_{\bf p}} (-i)^n
(p^0)^n \exp \left( i {\bf p} ({\bf x}'-{\bf x}) \right)
P(i p^0, -i p_k, \stackrel{\rightarrow}{\partial_0},
\stackrel{\rightarrow}{\partial_k})
\varphi({\bf x}, x_0) \\ & = & \int d^3 {\bf x} \frac{d^4 p}{(2
\pi)^3} \frac{\delta(p^0 - \omega_{\bf p})}{2 \omega_{\bf p}} (-
i)^n (\omega_{\bf p})^n e^{i {\bf p} ({\bf x}'-{\bf x})}
P(i p^0, -i p_k, \stackrel{\rightarrow}{\partial_0},
\stackrel{\rightarrow}{\partial_k})
\varphi({\bf x}, x_0) \\ & = & \int d^3 {\bf x} \frac{d^4 p}{(2
\pi)^3} \frac{\delta(p^0 - \omega_{\bf p})}{2 \omega_{\bf p}} (-
i)^n \sqrt{m^2 - (\nabla')^2}^n e^{i {\bf p}
({\bf x}'-{\bf x})} P(i p^0, -i p_k,
\stackrel{\rightarrow}{\partial_0},
\stackrel{\rightarrow}{\partial_k})
\varphi({\bf x}, x_0) \\
& = & ( -i )^n \sqrt{m^2 - (\nabla')^2 }^n \varphi ({\bf x}', x_0).
\end{eqnarray*}
Therefore we conclude that $\varphi$ satisfies the positive square root
of the Klein-Gordon equation.
\end{appendix}


\newpage
\begin{thebibliography}{9}
\bibitem{Ha92} J.J. Halliwell and M.E. Ortiz, Phys.~Rev.~D  {\bf
48} (1993) 748. \\
J.J. Halliwell and M.E. Ortiz, preprint CTP \# 2235, to appear in
the Proceedings of Journ\'ees Relativistes 93, World Scientific,
1993.
\bibitem{Ha88} J.J. Halliwell, Phys.~Rev.~D {\bf 38} (1988) 2468.
\bibitem{Te82} C. Teitelboim, Phys.~Rev.~D {\bf 25} (1982) 3159.
\bibitem{He92} M. Henneaux and C.
Teitelboim, Quantisation of Gauge Systems, (Princeton Univ. Press,
1992). \\
M. Henneaux, Classical Foundations of BRST
Symmetry, (Bibliopolis, Napoli, 1988). \\
M. Henneaux, Phys.~Rep.~{\bf 126} (1985) 1. \\
K. Sundermeyer, Constraint Dynamics, Lecture Notes
in Physics {\bf 169}, (Springer, Berlin, 1982).
\bibitem{Fe50} R.P. Feynman, Phys.~Rev.~{\bf 80} (1950) 440. \\
V. Fock, Physik.~Zeits.~Sowjetunion {\bf 12}
(1937) 404.
\bibitem{Fr75} E.S. Fradkin and G.A. Vilkovisky, Phys.~Lett.~B
{\bf 55} (1975) 224. \\
I.A. Batalin and G.A. Vilkovisky, Phys.~Lett.~B {\bf 69} (1977)
309. \\ E.S. Fradkin  and T.E. Fradkina, Phys.~Lett.~B {\bf 72}
(1978) 343.
\bibitem{Au85} A. Auerbach and S. Kivelson, Nucl.~Phys.~B {\bf
257} (1985) 799.
\bibitem{An93} A. Anderson, preprint, Imperial-TP-92-93-46.
\end{thebibliography}
\end{document}